
\input phyzzx

\overfullrule=0pt
\hsize=6.3truein
\vsize=9.0truein
\voffset=-0.1truein

\def\twod{two-dimensional}
\def\nab{{\scriptstyle \nabla}}
\hyphenation{Schwarz-schild}

\rightline{SU-ITP-93-10}
\rightline{hepth/9305030}
\rightline{May 1993}

\bigskip\bigskip
\title{Tachyon Hair on Two-Dimensional Black Holes\foot{Work
supported in part by NSF grant PHY-163D368.}}

\vfill
\author{Amanda Peet\foot{e-mail: peet@dormouse.stanford.edu}, Leonard
Susskind and L\'arus Thorlacius\foot{e-mail:
larus@dormouse.stanford.edu}}
\address{ Department of Physics \break Stanford University \break
            Stanford, CA 94305}
\vfill

\abstract
\singlespace
Static black holes in two-dimensional string theory can carry tachyon
hair.  Configurations which are non-singular at the event horizon
have non-vanishing asymptotic energy density.  Such solutions can be
smoothly extended through the event horizon and have non-vanishing
energy flux emerging from the past singularity.  Dynamical processes
will not change the amount of tachyon hair on a black hole.  In
particular, there will be no tachyon hair on a black hole formed in
gravitational collapse if the initial geometry is the linear dilaton
vacuum.  There also exist static solutions with finite total energy,
which have singular event horizons.  Simple dynamical arguments
suggest that black holes formed in gravitational collapse will not
have tachyon hair of this type.
\vfill\endpage

\REF\nohair{V.~Ginzburg and L.~Ozernoi
    \journal Soviet Phys., JETP & 20 (65) 689;  \hfil\break
     A.~Doroshkevich, Ya.~Zeldovich and I.~Novikov
    \journal Soviet Phys., JETP & 22 (66) 122;  \hfil\break
     W.~Israel
    \journal Phys. Rev. & 164 (67) 1776;
    \journal Comm. Math. Phys. & 8 (68) 245;  \hfil\break
     B.~Carter
    \journal Phys. Rev. Lett. & 26 (71) 331;  \hfil\break
     J.~D.~Bekenstein
    \journal Phys. Rev. & D5 (72) 1239;  \hfil\break
     J.~B.~Hartle, in {\it Magic without Magic,}
     ed. J.~Klauder (Freeman, 1972);  \hfil\break
     C.~Teitelboim
    \journal Phys. Rev. & D5 (72) 2941;  \hfil\break
     R.~H.~Price
    \journal Phys. Rev. & D5 (72) 2419: {\it ibid.} 2439.}
\REF\mueller{M.~Mueller
    \journal Nucl. Phys. & B337 (90) 37;  \hfil\break
     S.~Elitzur, A.~Forge and E.~Rabinovici
    \journal Nucl. Phys. & B359 (91) 581;  \hfil\break
     M.~Rocek, K.~Schoutens and A.~Sevrin
    \journal Phys. Lett. & B265 (91) 303;  \hfil\break
     G.~Mandal, A.~Sengupta and S.~R.~Wadia
    \journal Mod. Phys. Lett. & A6 (91) 1685.}
\REF\witten{E.~Witten
    \journal Phys. Rev. & D44 (91) 314.}
\REF\dalyk{S.~P.~de~Alwis and J.~Lykken
    \journal Phys. Lett. & B269 (91) 264.}
\REF\jorge{J.~G.~Russo, {\it Classical Solutions in Two-Dimensional
           String Theory and Gravitational Collapse,} University of
	   Texas preprint, UTTG-31-92, December 1992.}
\REF\chamin{S.~Chaudhuri and D.~Minic, {\it On the Black Hole
           Background of Two-Dimensional String Theory,} University
	   of Texas preprint, UTTG-30-92, December 1992.}
\REF\skrama{S.~K.~Rama, {\it Tachyon Back Reaction on $d=2$ Black
         Hole,} Trinity College Dublin preprint, TCD-3-93, March
         1993.}
\REF\ginsmoor{For a review of matrix models see P.~Ginsparg and
      G.~Moore, {\it Lectures on 2D Gravity and 2D String Theory},
      preprint YCTP-P23-92, LA-UR-92-~3479.}
\REF\cfmp{C.~G.~Callan, D.~Friedan, E.~J.~Martinec and M.~J.~Perry
    \journal Nucl. Phys. & B262 (85) 593;  \hfil\break
      E.~S.~Fradkin and A.~A.~Tseytlin
    \journal Nucl. Phys. & B271 (86) 561;  \hfil\break
      C.~G.~Callan and Z.~Gan
    \journal Nucl. Phys. & B272 (86) 647.}
\REF\tom{T.~Banks
    \journal Nucl. Phys. & B361 (91) 166;  \hfil\break
     A.~A.~Tseytlin
    \journal Phys. Lett. & B264 (91) 311.}
\REF\cst{A.~Cooper, L.~Susskind and L.~Thorlacius
    \journal Nucl. Phys. & B363 (91) 132; {\it Quantum Cosmology
     on the Worldsheet,} in {\it Proceedings of the ``Strings and
     Symmetries" workshop, Stony Brook, N.~Y., May 20-25, 1991,}
     W.~Siegel {\it et al.} eds., (World Scientific).}
\REF\das{S.~R.~Das
    \journal Mod. Phys. Lett. & A8 (93) 69.}
\REF\cghs{C.~G.~Callan, S.~B.~Giddings, J.~A.~Harvey and
     A.~Strominger
    \journal Phys. Rev. & D45 (92) R1005.}
\REF\dvv{R.~Dijkgraaf, E.~Verlinde and H.~Verlinde
    \journal Nucl. Phys. & B371 (92) 269.}
\REF\mswii{G.~Mandal, A.~Sengupta and S.~R.~Wadia
    \journal Mod. Phys. Lett. & A7 (92) 3703.}
\REF\jorgeii{J.~G.~Russo
    \journal Phys. Lett. & B300 (93) 336.}
\REF\lslt{L.~Susskind and L.~Thorlacius
    \journal Nucl. Phys. & B382 (92) 123.}
\REF\maoz{N.~Marcus and Y.~Oz, {\it The Spectrum of the 2D Black
     Hole, or Does the 2D black hole have tachyonic or W-hair?,}
     Tel-Aviv preprint, TAUP-2046-93, May 1993.}

\noindent I. \hfil\break
The classical ``no-hair theorems '' of general relativity severely
restrict the kinds of fields that can exist outside a black hole in
four spacetime dimensions [\nohair].  For example, in a static
geometry with a smooth event horizon the only fields which are well
behaved both asymptotically and on the event horizon are monopole
gravitational and electromagnetic fields.  Continuum string theory in
two spacetime dimensions has classical solutions which are analogs of
black holes [\mueller,\witten] and provides a simplified context in
which to study black hole physics.  It has been suggested that the
analog of the no-hair theorem for scalar fields does not hold for
these \twod\ geometries.  In particular, it was pointed out by Witten
[\witten] that the tachyon field of the \twod\ string theory has a
one-parameter family of solutions which are non-singular at the black
hole event horizon.  Subsequently, a number of authors have
considered the back-reaction of the tachyon field on the black hole
geometry using various approximation schemes [\dalyk-\skrama].
However, no consensus has been reached as to whether the
back-reaction destabilizes the configurations with tachyon hair and
leads to singularities at the event horizon.

In this note we will examine the issue of tachyon hair by
investigating solutions of the low-energy equations of motion of
\twod\ string theory with a non-vanishing tachyon background. We
define a black hole with tachyon hair to be a geometry with a smooth
event horizon and a tachyon field which is non-singular at that
horizon.  We find that solutions with hair exist, even when the
tachyon back-reaction on the geometry is taken into account, but that
they have non-vanishing asymptotic energy density and therefore
infinite ADM mass.  These static geometries with hair correspond to
eternal black holes where there is a balance between an outgoing and
an incoming energy flux. The outgoing flux emerges from the past
singularity and will not be present when a black hole is formed in
gravitational collapse.  We therefore conclude that dynamically
formed black holes cannot have tachyon hair of this type.\foot{One
can take the view that this type of tachyon hair defines
superselection sectors in the theory since the amount of hair cannot
be changed in dynamical processes.  In each sector the (infinite)
energy associated with the asymptotic tachyon field could then be
subtracted off to obtain a finite mass for a given configuration.}
When the tachyon back-reaction is included, there also exist
solutions with diverging string coupling and a curvature singularity
at the horizon~[\skrama].  We argue that such configurations will not
arise in dynamical evolution from non-singular initial data.

The issue of black hole hair comes up in attempts to describe black
hole geometries in the context of the $c=1$ matrix model [\ginsmoor].
 It is important to resolve whether one should look for a family of
black holes labelled by other parameters besides the mass.  At the
end we will comment on the possible relevance of our results to this
issue.

\noindent II. \hfil\break
Tree-level backgrounds in string theory are determined by the
conditions of Weyl invariance of the worldsheet sigma model.  These
can be obtained as the variational equations of an effective action
for the target space metric, dilaton and tachyon fields [\cfmp],
which has the following form in two
dimensions\foot{We take $\alpha'=2$ throughout.},
$$
S = {1\over 2\pi} \int d^2 x \sqrt{-g} \, e^{-2\phi} \,
        \Bigl\{R + 4\,(\nab \phi)^2 +  8
       - {1\over 2}\,(\nab T)^2 - V(T) + \cdots \Bigr\} \,,
\eqn\action
$$
The tachyon potential $V(T) = - T^2 + c T^3 + \ldots$ is non-linear
and only the leading term, which determines the tachyon mass, is
universal.  The higher order terms depend on the renormalization
prescription used to define the worldsheet theory [\tom,\cst] but our
conclusions will not be affected by their detailed form.  The
effective action receives corrections from worldsheet perturbation
theory, which are non-linear in the sigma model fields and of higher
order in derivatives.  We expect our qualitative conclusions to hold
when those corrections are included.  There will also be corrections
from string loops which contribute terms of higher order in the
string coupling $e^{\phi}$.  These will lead to interesting and
important quantum effects such as Hawking radiation.  In this work we
will only consider string theory at tree level.  The question of
whether or not classical \twod\ black holes can have tachyon hair is
relevant to the evolution of massive black holes in the quantum
theory.  One would, for example, expect the presence of black hole
hair to affect the Hawking temperature.

The classical equations of motion obtained from the action \action\
can be written,
$$\eqalign{
\nab ^2 T -2\nab \phi \cdot\nab T =&\> V '(T) \, ,  \cr
\nab ^2\phi - 2(\nab \phi)^2 =&\> -4 +{1\over 2}V(T) \, ,  \cr
2\nab _\mu \nab _\nu \phi - g_{\mu\nu} \nab ^2\phi =&\>
{1\over 2}\nab _\mu T\nab _\nu T
 -{1\over 4} g_{\mu\nu} (\nab T)^2 \, . \cr}
\eqn\coveom
$$
Note that the tachyon source terms in the equations for the metric
and dilaton are at least quadratic in $T$ while the tachyon equation
itself is linear in $T$ to leading order.  This means that one can
consistently view a weak tachyon field as a small perturbation and
ignore its back-reaction on the background geometry to leading order.
 We will come back to the issue of the back-reaction due to a finite
strength tachyon field later on.
Let us begin by looking for the static solutions of the
dilaton-gravity sector in the absence of tachyon matter.  They are
conveniently described using the \twod\ analog of Schwarzschild
coordinates, for which the metric has the form,
$$
ds^2 =\, -f(r)\,d t ^2 +\,{1\over f(r)}\,dr^2 \>.
\eqn\metric
$$
The equations in vacuum for $f$ and $\phi$ reduce to,
$$
\phi '' = \, 0
\quad ,\qquad
f '\phi '-2f\phi '^2 +4 = \, 0
\>.
\eqn\static
$$
The dilaton is a linear function of the spatial coordinate and any
constant part can be absorbed into a shift of $\phi$.
After rescaling the coordinates to make the metric approach the flat
Minkowski metric asymptotically we are left with a one-parameter
family of physically distinct static geometries,
$$
\phi (r) = -\sqrt{2}\,r  \quad ,\qquad
f(r) = 1-\mu e^{-2\sqrt{2}r}  \>.
\eqn\bhole
$$
All these solutions approach the linear dilaton vacuum in the
asymptotic region $r\rightarrow +\infty$.
The parameter $\mu$ is proportional to the mass of the black hole as
determined by the canonical ADM procedure, $\mu =\sqrt{2}\pi
M_{ADM}$.
For positive mass the geometry has a spacelike curvature singularity
inside an event horizon at
$r_h = {1\over 2\sqrt{2}} \log{({1\over \mu})}$
but for negative mass the singularity is naked.

We now wish to consider a weak static tachyon field outside a
positive mass black hole.  For this purpose it is convenient to
absorb a factor of the string coupling into the tachyon field,
$$
U = e^{-\phi} \, T  \>.
\eqn\redef
$$
Then the tachyon equation of motion becomes
$$
\nab ^2 U + \bigl(\nab ^2\phi - (\nab \phi)^2\bigr)\,U
=\, e^{-\phi}\,V '(e^\phi U) \>.
\eqn\lineq
$$
The redefined tachyon field has a standard kinetic term and it is $U$
which is the physical propagating mode in the system.  The mass term
for $U$ depends on the background geometry.  It reduces to zero in
the linear dilaton vacuum, as expected, and its effect therefore
vanishes in the asymptotic region.

The linearized equation for a static $U$ field on a black hole
background in our Schwarzschild coordinates is
$$
(1-\mu e^{-2\sqrt{2}r})\,U ''+2\sqrt{2}\mu e^{-2\sqrt{2}r}\,U '
-2\mu e^{-2\sqrt{2}r}\,U =\>0 \>.
\eqn\ueq
$$
It is straightforward to find the general solution for $U$ outside
the black hole event horizon [\das].  The change of variables $y=\mu
e^{-2\sqrt{2}r}$ takes \ueq\ to a hypergeometric equation,
$$
y(1{-}y)\,{d^2U\over dy^2} + (1{-}2y)\,{dU\over dy} -{1\over 4}\,U
=\>0 \>,
\eqn\hyper
$$
with regular singular points at $y=0$ and $y=1$.  The asymptotic
region of the black hole geometry is obtained as $y\rightarrow 0^+$
and the event horizon is approached as $y\rightarrow 1^-$.  The
general solution of \hyper\ is
$$
U(y) =\, A_1 \, u_1(y) + \, A_2 \, u_2(y) \>,
\eqn\ugeneral
$$
where the basic solutions are given in terms of a single
hypergeometric function,
$$
u_1(y) = \, F({1\over 2},{1\over 2};1;y)  \quad ,\qquad
u_2(y) = \, F({1\over 2},{1\over 2};1;1{-}y)  \>.
\eqn\ubasic
$$
The hypergeometric function is analytic in the interval $0<y<1$,
which corresponds to the black hole exterior, and has the following
asymptotic behavior near the endpoints,
$$\eqalign{
F({1\over 2},{1\over 2};1;y) &\sim
1 + {1\over 4} y + O(y^2) \qquad {\rm as}\quad
y\rightarrow 0 \>,   \cr
F({1\over 2},{1\over 2};1;1{-}y) &\sim
-{1\over \pi}\log{({y\over 4})}
\bigl( 1 + {1\over 4} y + O(y^2) \bigr)
-{1\over 2\pi} y + O(y^2)
\>\> {\rm as}\quad
y\rightarrow 0 \>,   \cr}
\eqn\ulimits
$$
Any solution with a non-vanishing coefficient of $u_1$ in \ugeneral\
has a logarithmic singularity at the event horizon.  Tachyon hair
must therefore be a pure $u_2$ configuration which is smooth at $y=1$
but behaves logarithmically for small $y$.  Going back to the
Schwarzschild coordinates we see that the hair is linear far away
from the black hole,
$$
U(r) \sim \beta \,r \qquad {\rm as}\quad  r\rightarrow\infty \>.
\eqn\linhair
$$
Since the mass of $U$ goes to zero in the asymptotic region it is
reasonable that static solutions approach linear behavior there.
Because of the non-vanishing gradient of $U$ there is a finite energy
density in the asymptotic region, ${\cal E} = {\beta^2\over 2\pi}$,
which means that the total ADM mass of a \twod\ black hole with
tachyon hair is divergent.  One might expect this infinite energy to
lead to a disastrous back-reaction on the geometry but this does not
happen because the coupling strength $e^\phi$ goes to zero
asymptotically, due to the linear dilaton, and the matter decouples
from the gravitational sector.  Because we are working in two
spacetime dimensions such solutions can be given an interpretation in
terms of balancing incoming and outgoing energy fluxes carried by
(zero-momentum) tachyons.  To see this, write the linear $U$ field in
terms of null coordinates, $U(r^\pm)={\beta\over 2}(r^+{-}r^-)$, and
observe that $T_{++}=T_{--}={\beta^2\over 4\pi}$.  Usually one
associates a non-vanishing energy flux with a propagating field, and
one can obtain the above linear $U$ field as the infinite wavelength
limit of such an oscillating mode.

A solution built out of $u_2$ alone is perfectly smooth at the event
horizon and can be extended through it into the black hole interior.
This can be seen by making a coordinate transformation to
the two-dimensional analog of Kruskal-Szekeres coordinates, $U,V$, in
which the black hole geometry is manifestly regular at the event
horizon,
$$
ds^2 =\, {dU\,dV \over 2 U V{-}\mu} \>.
\eqn\kruskal
$$
A static geometry will only depend on the product $\xi = -2U V$.  The
relationship between $\xi$ and our spatial variable $y$ is
$$
y={1\over 1+{\xi\over \mu}} \>,
\eqn\yxi
$$
and it is non-singular at the event horizon.  This means that a
solution which is regular as $y\rightarrow 1$ will also be regular as
a function of
$\xi\sim 0$ and can be extended into the black hole interior to
negative values of $\xi$.  This indicates that a flux interpretation
of the solutions with tachyon hair involves fluxes which pass through
the event horizon and terminate (originate) on the future (past)
singularity.  If, on the other hand, the black hole is formed
dynamically by gravitational collapse there is no past singularity in
the geometry and the tachyon hair would have to be supported by an
outgoing flux emerging from the original vacuum.\foot{Note that this
outgoing flux would be there at the classical level and has nothing
to do with the quantum mechanical Hawking flux which emanates from a
black hole.}  This would require imposing non-standard boundary
conditions on the system at an early time, before the collapse takes
place, and we conclude that dynamically formed black holes cannot
carry tachyon hair of this type.

The ability of eternal black holes in \twod\ string theory to carry
tachyon hair relies on the particular manner in which the tachyon
field is coupled to the gravitational theory.  Consider for
comparison the model suggested by Callan {\it et~al.} [\cghs] for the
study of black hole physics in two dimensions.  The classical gravity
sector in this theory is identical to the leading terms in the \twod\
string theory \action\ so in the absence of matter it has the same
classical black hole solutions \bhole .  The matter sector consists
of a set of conformally coupled scalar fields which replace the
tachyon of the string theory.  In Schwarzschild coordinates the
classical equation of motion of a conformal scalar field, $a$, is
$$
f a '' + f ' a ' = \,0 \>,
\eqn\confeq
$$
and the general solution in a black hole background is given by
$$
a(r) = \, a_0 + a_1 \log{(e^{2\sqrt{2}\,r}-\mu)} \> .
\eqn\asol
$$
The only solution which is regular at the event horizon is a constant
which means that black holes cannot have non-trivial scalar hair in
this \twod\ model.

\noindent III. \hfil\break
Will our conclusions be qualitatively changed by higher order
corrections to the effective action \action ?  Let us first consider
contributions from higher orders in sigma model perturbation theory.
The static black hole solution \bhole\ is only valid to lowest order
but it has been suggested that an exact solution to the Weyl
invariance conditions on the metric and dilaton, in the absence of a
tachyon background, is provided by a deformed ``cigar''
geometry~[\dvv], for which the detailed spatial dependence of $\phi$
and $f$ differs from \bhole\ but there is still a smooth horizon and
the fields approach the linear dilaton vacuum asymptotically.

It is easy to see that near a smooth event horizon the linearized
scalar field equation always has one non-singular solution and one
which diverges logarithmically.  Asymptotically far away the mass
term goes to zero and the general solution is a linear function of
the Schwarzschild coordinate.  For the case of a lowest order black
hole background \bhole\ we were able to interpolate between the event
horizon and asymptotic region with the exact solution \ugeneral\ and
we found that solutions which are smooth at the event horizon are in
fact orthogonal to those that are well behaved asymptotically.  While
this exact orthogonality will presumably not survive to higher orders
it is clear that the corrections will not generically change the fact
that solutions, which are smooth at the event horizon, will have a
non-vanishing energy density at infinity.

What about the back-reaction of the tachyon field on the geometry?
S.~K.~Rama has argued that the effect of the back-reaction is to
destabilize any smooth solution at the horizon [\skrama].  This would
of course rule out the existence of any tachyon hair in the full
theory, even if one is willing to work with solutions which have
non-vanishing asymptotic energy density.  However, the analysis of
reference [\skrama] is incomplete.  While one can indeed find static
solutions of \coveom\ with non-vanishing $T$, and a curvature
singularity along with diverging string coupling at the event
horizon, there also exist configurations with non-vanishing tachyon
hair which have a perfectly smooth event horizon.  Let us again work
in a Schwarzschild coordinate system \metric\ and consider the full
set of static equations,
$$\eqalign{
fT ''+f 'T '-2f\phi 'T ' =&\, V '(T)  \>,  \cr
f\phi ''+f '\phi '-2f\phi '^2 =&\, {-}4 +{1\over 2} V(T)  \>,  \cr
\phi '' =&\, {1\over 4} T '^2  \>.  \cr}
\eqn\stateqs
$$
Let $x=r{-}r_h$ be the coordinate distance from the event horizon and
look for a regular solution of the form,
$$\eqalign{
T(x) =&\, \sum_{k=0}^\infty t_k \,x^k  \>,  \cr
\phi(x) =&\, \sum_{k=0}^\infty \phi_k \,x^k  \>,  \cr
f(x) =&\, \sum_{k=1}^\infty f_k \,x^k  \>,  \cr}
\eqn\ansatz
$$
which is expected to be valid near the event horizon.
The power series coefficients can be determined by inserting \ansatz\
into \stateqs\ and working order by order in $x$.  We choose to
parametrize the family of solutions by the tachyon and dilaton fields
at the event horizon and we use the freedom to rescale the coordinate
$x$ to fix the ratio between $f_1$ and $f_2$ as $-1$.  The leading
terms in the solution are then given by,
$$\eqalign{
T(x) =&\, t_0 +{V '(t_0)\over 4{-}{1\over 2}V(t_0)}\> x
    +{{1\over 4}V '(t_0)V ''(t_0)\over (4{-}{1\over 2}
    V(t_0))^2}\> x^2
    + \ldots  \>,  \cr
\phi(x) =&\, \phi_0 - x
    +{{1\over 8}V '(t_0)^2\over (4{-}{1\over 2}V(t_0))^2}\> x^2
    + \ldots  \>,  \cr
f(x) =&\, \bigl(4{-}{1\over 2}V(t_0)\bigr)(x-x^2) + \ldots  \>.  \cr}
\eqn\terms
$$
When $T=0$ this gives us the leading terms in the expansion of the
black hole solution \bhole\ (up to a rescaling of the coordinate by
$\sqrt 2$).  The solution will behave in some complicated way as we
move away from the horizon but the asymptotic region is governed by
the linear dilaton and the solution will match onto a linear $U$
field there.  Thus there exist infinite energy configurations with
tachyon hair which have a non-singular event horizon.

There are other solutions, where the tachyon back-reaction leads to a
singular event horizon.  The dominant small $x$ behavior can be
parametrized as follows,
$$\eqalign{
T(x) =&\, \eta \log{x}  \>,  \cr
\phi(x) =&\, \phi_0 - {\eta^2\over 4}\log{x}  \>,  \cr
f(x) =&\, x^{1{-}{\eta^2\over 2}}  \>.  \cr}
\eqn\blowup
$$
The scalar curvature blows up as the horizon is approached,
$R\sim  {\eta^2\over 2}(1{-}{\eta^2\over 2})\,x^{-1{-}{\eta^2\over
2}}$,
and the string coupling also diverges there,
$e^\phi \sim e^{\phi_0}\> x^{-{\eta^2\over 2}}$.

These singular solutions presumably correspond to the ones discussed
in reference [\skrama].  Can they be the end product when a \twod\
black hole is formed in gravitational collapse?  Consider a set of
initial data describing infalling tachyon matter and assume that
after some time a black hole forms with a spacelike curvature
singularity inside an apparent horizon.  A singular horizon might
develop from an instability in the evolution of the fields at or near
the apparent horizon, which would lead to a new curvature singularity
separate from the one in the black hole interior.  This scenario can
be ruled out using the classical equations of motion.  For this we
adapt a conformal gauge argument previously given [\lslt] in the
context of the dilaton gravity model of Callan {\it et al.} [\cghs].
Let $\sigma^\pm$ be a set of null conformal coordinates;
$ds^2 = -e^{2\rho (\sigma)}\,d\sigma^+\,d\sigma^-$.
In this gauge the second equation in \coveom\ becomes
$$
\partial_+\partial_-\phi
-2\partial_+\phi\partial_-\phi
= \,e^{2\rho}\,\bigl(
1-{1\over 8}V(T)\bigr) \>.
\eqn\dileq
$$
The incoming matter is initially described by some left-moving
tachyon field in the asymptotic weak-coupling region and we'll assume
that there is a null line, $\sigma^+=\sigma^+_0$, below which we have
the vacuum configuration (see figure 1a).  On this null line the
dilaton field is a monotonic function of $\sigma^-$, increasing as we
go deeper into the strong coupling region.  The dilaton diverges at
the event horizon in the singular static solution \blowup .  In order
for such a blow-up to occur, away from the strong-coupling
singularity in the black hole interior, the dilaton must first
develop a local maximum, which then has to grow without bound.  But
if the dilaton field is a monotonic function of $\sigma^-$ on some
null line of constant $\sigma^+=\sigma^+_0$ then the equation of
motion \dileq\ ensures that it will remain monotonic on all null
lines $\sigma^+ > \sigma^+_0$.  If a local maximum were to occur it
would have to be accompanied by a local minimum closer to the strong
coupling region, as illustrated in figure 1b.  Call the points where
the extrema occur $\sigma^-_1(\sigma^+)$ and $\sigma^-_2(\sigma^+)$.
At these points $\partial_-\phi$ vanishes and \dileq\ reduces to
$$
\partial_+(\partial_-\phi)
= \,e^{2\rho}\,\bigl(
1-{1\over 8}V(T)\bigr) \>.
\eqn\dilcond
$$
As long as the right hand side of \dilcond\ remains positive the two
curves $\sigma^-_1(\sigma^+)$ and $\sigma^-_2(\sigma^+)$ must
approach each other as $\sigma^+$ increases.  The dilaton field is
monotonic on $\sigma^+=\sigma^+_0$ so there would have to be a
smallest value of $\sigma^+$ where the extrema first occur and at
which $\sigma^-_1=\sigma^-_2$.  But, since the equations of motion
prevent $\sigma^-_1$ and $\sigma^-_2$ from separating it follows that
the dilaton must remain monotonic for all $\sigma^+ > \sigma^+_0$.
This conclusion is in agreement with the results of Russo [\jorge]
who studied black hole formation due to incoming tachyon matter and
found no evidence for singular horizons.

Our argument would fail if the shape of the tachyon potential were
such that $V(T)>8$ were possible for some value of $T$.  This,
however, is very unlikely since in order to become positive the
tachyon potential would have to have a minimum somewhere and that
would correspond to a stable vacuum of the bosonic string.\foot{The
tachyon potential has been computed to all orders in $T$ and found to
be negative definite [\cst].  The form of the potential is of course
highly sensitive to the sigma model renormalization prescription
used, but we expect the feature that the potential does not turn
around and become positive to be universal.}

\noindent IV. \hfil\break
We have found that non-singular tachyon hair is a feature of
eternal static black holes only and does not arise when the black
hole is dynamically formed.  We have also seen that new
singularities,
associated with non-vanishing tachyon hair of the type \blowup ,
cannot be formed at the horizon.  We have not given proof that the
original black hole singularity cannot merge with the apparent
horizon to form a singular event horizon, which might carry singular
tachyon hair.  However, the nature of the curvature singularity at
the event horizon in \blowup\ appears to be quite different from that
of the singularity inside a classical black hole \bhole , where the
curvature blows up at the same rate as the coupling strength, $R =
8\mu e^{2\phi}$.  We find it more likely that the tachyon field will
be absorbed by the black hole or radiated away as the geometry
settles down to its final state.

The asymptotic linear $U$ field of a \twod\ black hole with tachyon
hair can be given an interpretation in terms of the fermion fields of
the $c=1$ matrix model.  The standard bosonization formulas identify
$\partial_rU$ with a shift in the fermion number density.
Non-singular tachyon hair of the type discussed above then
corresponds to shifting the fermi level in the inverted harmonic
oscillator potential of the matrix model.  On the other hand, recent
efforts to understand black holes in the context of the matrix model
suggest that the height of the fermi level corresponds to the black
hole mass [\das,\mswii,\jorgeii].  These two interpretations appear
to be in conflict, unless the mass and tachyon hair of static black
holes in string theory are correlated and there is only a one
parameter family of physically distinct solutions.

In conclusion we would like to emphasize once again that \twod\
tachyon hair is very different from the hair envisioned in
four-dimensional no-hair conjectures.  Usually black hole hair refers
to fields which might result from the collapse of matter to form the
black hole whereas we find that tachyon hair on two-dimensional black
holes is infinitely long and can neither be grown nor cut.

\noindent
\undertext{Note added}: While preparing this manuscript we became
aware of the work of N.~Marcus and Y.~Oz [\maoz] which has some
overlap with the present work.

\refout
\end